\documentclass{PoS}

\title{Determination of the mass anomalous dimension for $N_f=12$ and $N_f=9$ SU($3$) gauge theories}

\ShortTitle{Determination of the mass anomalous dimension for $N_f=12$ and $N_f=9$ SU($3$) gauge theories}

\author{\speaker{Etsuko Itou}\\
        High Energy Accelerator Research Organization (KEK), Tsukuba 305-0801, Japan\\\
        E-mail: \email{eitou@post.kek.jp}}

\author{Akio Tomiya\\
        Department of Physics, Osaka University, Toyonaka 560-0043, Japan \\
        E-mail: \email{akio@het.phys.sci.osaka-u.ac.jp}}

\abstract{We show the numerical simulation result for the mass anomalous dimension of the SU($3$) gauge theory coupled to $N_f = 12$ fundamental fermions.
We use two independent methods, namely the step scaling method and the hyperscaling method of the Dirac mode number, to determine the anomalous dimension in the vicinity of the infrared fixed point of the theory.
We show the continuum extrapolations keeping the renormalized coupling constant as a reference in both analyses.

Furthermore, some recent works seems to suggest the lower boundary of the conformal window of the SU($3$) gauge theory exists between $N_f=8$ and $10$.
We also briefly report our new project, in which the numerical simulation of the SU($3$) gauge theory coupled to $N_f=9$ fundamental fermions has been performed.}

\FullConference{The 32nd International Symposium on Lattice Field Theory,\\
		23-28 June, 2014\\
		Columbia University New York, NY}

\usepackage{amsmath}
\newcommand{\beq}{\begin{eqnarray}}
\newcommand{\eeq}{\end{eqnarray}}

\begin{document}

\section{Introduction}
Critical exponent at the conformal fixed point is one of the most important objects in quantum field theories.
Recent studies (see references in Refs.~\cite{Kuti}) reveal the existence of the infrared fixed point (IRFP) of several many flavor non-abelian gauge theories.
We focus on the SU($3$) gauge theory coupled to $N_f=12$ massless fermions.
In our previous works, we found the IRFP using the step scaling method in the Twisted Polyakov Loop (TPL) scheme~\cite{TPL} and obtained the preliminary result of the mass anomalous dimension around the IRFP~\cite{ano-dim}.
In this proceedings, we show the updated data including a new large lattice volume data.
We obtain the value of the mass anomalous dimension around the IRFP based on two independent methods, namely the step scaling~\cite{Luscher:1991wu,Capitani:1998mq} and the hyperscaling method of the mode number~\cite{DelDebbio:2010ze,Patella:2012da,Cheng:2013eu}.

Furthermore, some recent works seems to suggest the lower boundary of the conformal window of the SU($3$) gauge theory exists between $N_f=8$ and $10$.
We also report that our new project for the SU($3$) gauge theory coupled to $N_f=9$ fundamental fermions to determine the range of the conformal window.

\section{Mass step scaling function for SU($3$) $N_f=12$ theory}
The step  scaling is one of established methods to obtain the renormalization factors, {\it e.g.} $Z_g$ and $Z_m$, using lattice simulations~\cite{Luscher:1991wu,Capitani:1998mq}.
We observe the growth ratio of the renormalized quantity when the lattice extent becomes $s$ times larger with fixed value of bare coupling constant ($\beta$). Here the scale "s" is the step scaling parameter.
According to the PCAC relation, the mass renormalization factor is inversely related with the renormalization factor of the pseudo scalar operator ($Z_P$), so that we can obtain it even for the massless fermions by the measurement of the renormalization factor of the pseudo scalar operator.

We use the renormalization scheme of the pseudo scalar operator proposed by Ref.~\cite{ano-dim} .
The renormalization factor is defined by the ratio of the pseudo scalar correlators between the nonperturbative and tree level ones at the fixed propagation length,
\beq
Z_P= \sqrt{ \frac{C_P^{\mathrm{tree}}(t)}{C_P (t)} } \mbox{ at fixed t},
\eeq
where $C_{H}(t) = \sum_{\vec{x}} \langle P(t,\vec{x}) P(0,\vec{0}) \rangle$ and $P(t,\vec{x})$ denotes the pseudo scalar operator.

On the lattice, this scheme has two independent scales, the propagation time ($t$) and lattice temporal size ($T$).
Here, we always fix the ratio of $T$ and $L$, and $L$ is identified with the renormalization scale ($\mu=1/L$).
To obtain the factor $Z_P$, we fix the ratio $r=t/T$, where $r$ takes a value $0< r \le1/2$ because of the periodic boundary condition on the lattice. 
Practically, the larger $r$ reduces the discretization error. We take $r=1/2$ in this analysis.

To obtain the scale dependence of the renormalization factor of the operator, we measure the mass step scaling function given by
\beq
\Sigma_P (\beta , a/L; s)&=&  \frac{Z_P (\beta, a/sL)}{Z_P(\beta, a/L)}, \label{eq:disc-sigma}
\eeq
using the lattice simulation.
The mass step scaling function on the lattice includes the discretization error.
To remove it, we take the continuum limit ($a \rightarrow 0$) keeping the renormalized coupling ($u=g_R^2 (1/L)$) constant;
\beq
\sigma_P (u,s) &=& \left. \lim_{a \rightarrow 0} \Sigma_P (\beta, a/L;s) \right|_{u=const}.\label{eq:def-sigma}
\eeq

In the continuum limit, this mass step scaling function is related to the mass anomalous dimension. 
This relation becomes simple when the theory is conformal, and we can estimate the anomalous dimension at the fixed point with the following equation:
\beq
\gamma_m^*(u^*)=-\frac{\log |\sigma_P (u^*,s)|}{\log|s|}.
\eeq
Here $u^*$ denotes the fixed point coupling constant.

Our numerical simulation has been carried out using the following lattice setup.
The gauge configurations are generated by the Hybrid Monte Carlo algorithm, and we use the Wilson gauge and the naive staggered fermion actions.
We introduce the twisted boundary conditions for $x, y$ directions and impose the usual periodic boundary condition for $z, t$ directions, which is the same setup with our previous work~\cite{TPL}.
Because of the twisted boundary conditions the fermion determinant is regularized even in the massless case, so that we carry out an exact massless simulation to generate these configurations.
The simulations are carried out with several lattice sizes ($L/a=8,10,12,16$, $20$ and $24$)
at the fixed point of the renormalized gauge coupling in the TPL scheme~\cite{TPL}.

In the paper~\cite{TPL}, the IRFP is found at 
\beq
g_{\mathrm{TPL}}^{*2} = 2.69 \pm 0.14 (\mbox{stat.}) ^{+0}_{-0.16} (\mbox{syst.}), 
\eeq
in the TPL scheme.
We use the tuned value of $\beta$ where the TPL coupling is the fixed point value for each $(L/a)^4$ lattices as shown in Table~\ref{table:beta-L}.
We generate the configurations using these parameters on $(L/a)^3 \times 2L/a$ lattices and neglect the possibility of induced scale violation coming from the change of lattice volume $(L/a)^4 \rightarrow 2(L/a)^4$ since we carefully take the continuum limit.
We measure the pseudo scalar correlator for $30,000$--$80,000$ trajectories for each $(\beta, L/a)$ combination.
\begin{table}[h]
\begin{center}
\begin{tabular}{|c|c|c|c|c|}
\hline
{} & {} &$g_{\mathrm{TPL}}^{2}=2.475$  & $g_{\mathrm{TPL}}^{2}=2.686$ & $g_{\mathrm{TPL}}^{2}=2.823$ \\     
\hline \hline
L/a & T/a & $\beta$ & $\beta$ & $\beta$  \\  
\hline
8   &  16 & 5.796          &  5.414           & 5.181              \\
10   &  20 &  5.998          &  5.653           & 5.450            \\
12   &  24 &  6.121          &  5.786           & 5.588              \\
\hline
\end{tabular}
\caption{ The values of $\beta$ for each $L/a$ which give the TPL coupling constant at the IRFP.} \label{table:beta-L}
\end{center}
\end{table}

\begin{figure}[h]
\begin{center}
\includegraphics[scale=0.4]{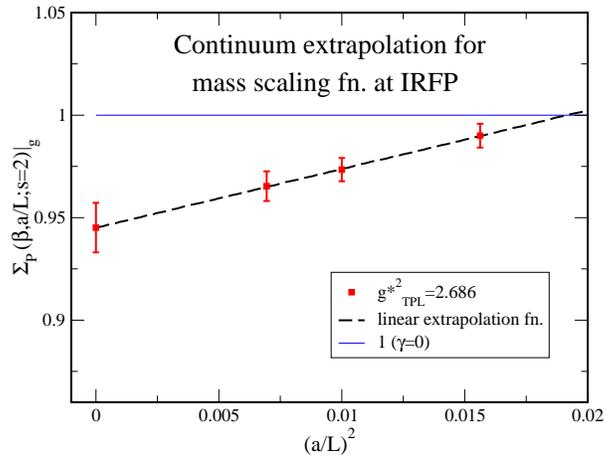}
\caption{Continuum extrapolation for the mass step scaling function at $g^{*2}_{\mathrm{TPL}}=2.686$.}
\label{fig:stepscaling}
\end{center}
\end{figure}

Figure~\ref{fig:stepscaling} shows the continuum extrapolation of the mass step scaling function ($\Sigma_P (\beta, a/L; s=2)$).
We take the three-point linear extrapolation in $(a/L)^2$, which is drawn in black dashed.
We take the result for $u=2.686$ as a fixed point coupling as a central result.
The mass anomalous dimension of the central result is given 
$\gamma_m^*= 0.081 \hspace{3pt} \pm 0.018 (\mbox{stat.}) $.
The error denotes the statistical one, which is estimated by the bootstrap method.

We also estimate the systematic uncertainties by considering the uncertainty of $g_{\mathrm{TPL}}^{*2}$.
Finally, we obtain the mass anomalous dimension 
\beq
\gamma_m^*&=& 0.081 \hspace{5pt} {\pm0.018} (\mbox{stat.})  \hspace{5pt}_{-0}^{+0.025} (\mbox{syst.}). \mbox{ (preliminary)} \label{eq:step-gamma}
\eeq
The detailed analysis will appear in the forthcoming paper.

\section{Hyperscaling for the mode number of $N_f=12$ theory}
The other promising method to observe the mass anomalous dimension at the IRFP is based on the hyperscaling of the spectral density ($\rho (\omega)$) as a function of the eigenvalue of the (massless) Dirac operator ($\omega$) given in Refs.~\cite{DelDebbio:2010ze,Patella:2012da,Cheng:2013eu}.
The mode number is the better quantity to observe the hyperscaling, since it is a renormalization group invariant.
 The mode number in the unit volume, whose absolute value of the Dirac eigenvalues is lower than a cutoff ($\lambda$), is given by
 \beq
 \nu (\lambda) = 2 \int_0^\lambda \rho(\omega) d \omega.
 \eeq
 
 At the IRFP in the infinite volume and continuum limits, there is the hyperscaling of the Dirac mode number,
 \beq
 \nu (\lambda) = C_0 \lambda ^{\frac{4}{1+ \gamma_m^*}},
 \eeq
for whole region of $\lambda$.
Here the exponent $\gamma_m^*$ is the mass anomalous dimension at the IRFP.
 On the finite lattice, both low and high eigenmodes suffer from the finite volume effects and the lattice cutoff ($a$) respectively.
We expect that there is a scaling region of $\lambda$ showing the hyperscaling in a middle range of $\lambda$.

Furthermore, there is a scale invariance in the scaling region at IRFP.
Our lattices shown in Table~\ref{table:beta-L} consist the fixed point coupling constant, namely $g^{*2}_{\mathrm{TPL}} (L)= g^{*2}_{\mathrm{TPL}} (sL)$, in the continuum limit.
Let us consider the scale transformation $a \rightarrow a' = a/s$. 
If the theory is conformal, then we expect that the shape of $\nu(\lambda)$ in the scaling region would indicate the invariance under the scale transformation,
\beq
a \lambda \rightarrow a \lambda' = s^{-(1+\gamma_m (\lambda))} a \lambda,  \label{eq:rescale-lambda}
\eeq
when the lattice setup approaches to the continuum limit.

Therefore, we expect two things:
\begin{enumerate}
\item there is a scaling region where the shape of $\nu (\lambda)$ is invariant under the scale transformation Eq.~(\ref{eq:rescale-lambda})
\item in the scaling region the value of the mass anomalous dimension is consistent with the result of the step scaling method, namely $\gamma_m^*$ in Eq.~(\ref{eq:step-gamma})
\end{enumerate}
 
To study these scaling properties, we measure the mode number whose eigenvalue is lower than $\lambda$ for the lattice sets shown in Table~\ref{table:beta-L2}. 
\begin{table}[h]
\begin{center}
\begin{tabular}{|c||c|c|c||c|c|c|}
\hline
 $\beta$ & $L/a$ & $T/a$ & $\Delta (a \lambda)$  & sL/a & sT/a & $ \Delta (a \lambda)$ \\  
\hline
 5.414     &8 & 16      & 0.02     & 16 & 32 & 0.01         \\
 5.653     &10 & 20      & 0.02   & 20 & 40 & 0.01         \\
 5.786     & 12 &24     & 0.02    & 24 & 48 & 0.01          \\
 \hline
\end{tabular}
\caption{ The values of $\beta$ for each $L/a$ which give the TPL coupling constant at the IRFP.} \label{table:beta-L2}
\end{center}
\end{table}
We use the stochastic method called the projector method~\cite{Giusti:2008vb}.
The number of configurations for each lattice setup is $30$ with $100$--$200$ separated Monte Carlo trajectories.
The computed mode number is in the range  $100 < \nu (\lambda) < 2,000$ with the tiny interval $\Delta (a \lambda) = 0.01$ or $0.02$.

\begin{figure}[h]
\vspace{0.5cm}
\begin{center}
\includegraphics[scale=0.5]{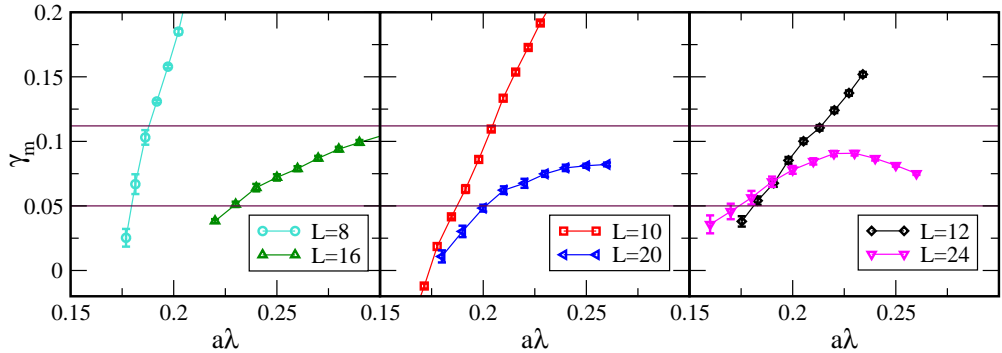}
\caption{Mass anomalous dimension from the hyperscaling analysis. The  region between two straight lines at $\gamma_m=0.050$ and $0.112$ denotes $1$-$\sigma$ bound of the result for the step scaling method.  }
\label{fig:hyperscaling}
\end{center}
\end{figure}

The preliminary results are shown in Fig.~\ref{fig:hyperscaling}.
The results for two lattice extents with the same $\beta$ are plotted in each panel.
Here for the smaller lattices ($(L/a)^3 \times (T/a)$), the horizontal axis is rescaled as Eq.~(\ref{eq:rescale-lambda}).
Each errorbar denotes only statistical error, which is estimated by the bootstrap method.
From the left to right panels, the continuum limit of the $\gamma_m$ are taken keeping the renormalized coupling constant.
We find that the tendency two lattice data close to each other from left to right.
Furthermore, the finest lattice sets ($12^3 \times 24$  and $24^3 \times 48$) overlap around $0.05 \le \gamma_m \le 0.08$. 
It is a signal of the scale invariance and the value of the mass anomalous dimension is consistent with the $1$-$\sigma$ bound of the step scaling result (Eq.~(\ref{eq:step-gamma})).
That is very promising.

The middle lattice sets ($10^3 \times 20$ and $20^3 \times 40$) also gives a signal of the scale invariance around $\gamma_m=0.02$, although the corseted lattice sets ($8^3 \times 16$ and $16^3 \times 32$) do not.
One of the reasons, that the coaster lattices do not show the scale invariance, is the smallness of the volume and the coarse lattice spacing.
The results of Ref.~\cite{Cheng:2013eu} show a clear the plateau of the anomalous dimension around the IRFP.
They use the improved staggered fermion action and the improvement would change the UV property.
Therefore the scaling region might be broader than our results.

\section{Status for the SU($3$) $N_f=9$ theory}
According to several recent studies, the boundary of the conformal window of the SU($3$) gauge theory coupled to the fundamental fermions would be expected around $8 \le N_f \le 10$.
We started the numerical simulation to study the IR behavior for $N_f=9$ case to determine the range of $N_f$ in the conformal window.

Our numerical simulation have been performed using the following setup. 
The gauge configurations are generated by the Rational Hybrid Monte Carlo (RHMC) algorithm with the Iwasaki gauge and the naive Wilson fermion actions. 
We use the $22$nd polynomial approximation with the interval $10^{-3} \le x \le 1$ where $x$ is the determinant of the Wilson fermion kernel for the RHMC algorithm.

The simulations have been carried out with lattice sizes $L/a = 8, 10, 12,14$ and $16$ with twice larger temporal lattice extent.
To obtain the line of the same renormalized mass in $\beta$--$\kappa$ plane, we generate $1, 000$ -- $5, 000$ trajectories for each parameter and measure the PCAC mass at several $\beta$ and $\kappa$.

\section{Summary} 
The mass anomalous dimension around the IRFP of the SU($3$) $N_f=12$ theory is obtained using the step scaling method,
\beq
\gamma_m^*&=& 0.081 \hspace{5pt} {\pm0.018} (\mbox{stat.})  \hspace{5pt}_{-0}^{+0.025} (\mbox{syst.}). \mbox{ (preliminary)}
\eeq
We found that the result of the hyperscaling method of the Dirac mode number also show the scale invariance and the value of the mass anomalous dimension in the scaling region is consistent with above value.
The detailed analysis will appear in the forthcoming paper.
On the other hand, the recent other results using the improved staggered action suggest the larger value of the mass anomalous dimensions $\gamma_m^* \approx 0.24$ \cite{Cheng:2013xha,Lombardo:2014pda}.
Although the difference is not so serious since it is roughly $2$-$\sigma$ consistent each other, it might be caused by the absence of the zero mode in our lattice setup because of the twisted boundary condition.
It might be worth investigating the nonperturbative effect of the zero mode to the universal quantity.

\section*{Acknowledgements}
We would like to thank S.~Aoki, A.~Hasenfratz and S.~M.~Nishigaki for several useful comments and discussions.
We also thank A.~Patella and M.~Honda for their helps to developing the numerical simulation code.
Numerical simulation was carried out on
Hitachi SR16000 at YITP, Kyoto University,
NEC SX-8R at RCNP, Osaka University,
and Hitachi SR16000 and IBM System Blue Gene Solution at KEK 
under its Large-Scale Simulation Program
(No.~13/14-20), as well as on the GPU cluster at Osaka University.
We acknowledge Japan Lattice Data Grid for data
transfer and storage.
E.I. is supported in part by
Strategic Programs for Innovative Research (SPIRE) Field 5.


\end{document}